\documentclass[preprintnumbers,amsmath,
amssymb,floatfix,8pt,prd,twocolumn,showpacs,
superscriptaddress,nofootinbib]{revtex4}
\usepackage{graphicx}
\usepackage{epsfig}
\usepackage{bm}
\usepackage{amsfonts}

\begin{document}

\title{  Mimetic Attractors }

\author{Muhammad Raza}
\email{mraza@zju.edu.cn, mreza06@gmail.com}
 \affiliation{Department of Mathematics, COMSATS Institute of Information Technology, Sahiwal 57000, Pakistan\\ and State Key Lab of Modern Optical Instrumentation,
Centre for Optical and Electromagnetic Research,
Department of Optical Engineering, Zhejiang University, Hangzhou 310058, China}

\author{ Kairat Myrzakulov}
\email{myrzakulov_kr@enu.kz}
\affiliation{Eurasian International Center for Theoretical Physics and Department of General \& Theoretical Physics, Eurasian National University, Astana 010008, Kazakhstan}

\author{ Davood Momeni}
\email{d.momeni@yahoo.com}
 \affiliation{Eurasian International Center for Theoretical Physics and Department of General \& Theoretical Physics, Eurasian National University, Astana 010008, Kazakhstan}

\author{ Ratbay Myrzakulov}
\email{rmyrzakulov@gmail.com}\affiliation{Eurasian International Center for Theoretical Physics and Department of General \& Theoretical Physics, Eurasian National University, Astana 010008, Kazakhstan}

\begin{abstract}
{\bf Abstract:}
 In this paper,we  investigate the mathematical modeling for the cosmological attractors propagated in mimetic gravity  upon which an interacting dark energy-dark matter is supposed to be existed.  The average value of the interaction of these percentages, namely $\Gamma_i$ say, may be used to investigate generally the modeling of an attractor; the actual value could only be determined by data in any particular case. We have seen, for example, that it was led to investigate the subject of initially invariant submanifolds.

\end{abstract}

 \maketitle

{\it Introduction}
In many passages of our studies on large scale structure of the whole Universe,  observational data  had distinctly stated that the current epoch of Universe which it dominated here give an explanation of the acceleration expansion of the Universe. Namely the data from supernovae SNe Ia {\cite{c1}}, cosmic microwave
background radiations via WMAP {\cite{c2}}, galaxy redshift surveys
via SDSS {\cite{c3}} and galactic X-ray {\cite{c4}}. But only the means of observing and connecting them together with theory; that the meaning of all phenomena, and the reason of their peculiar connexions, was a philosophical problem which required to be attacked from a different point of view; and that the significance especially which lay in the phenomena of  matter contents and exotic form of it would only unfold itself if by an exhaustive survey of the entire fully theoretical description, individually, consistently, and physically, we gain the necessary data for deciding what meaning attaches to the existence of this current acceleration expansion, or dark energy (DE) and dark matter (DM)  in the contents of the universe. DE is considered as an exotic matter with negative effective pressure to energy density  ratio in a very recent epoch $z\sim0.7$. The existence of such type of energy brings two important unsolved problems: The fine tuning and cosmic coincidence. Perhaps the best candidates for  DE are  the quintessence scalar field or a phantom energy field
\cite{Hossain:2014zma}.
The geometrical modifications of Einstein gravity are also interesting and have been studied widely in literatures \cite{Copeland:2006wr}-\cite{Capozziello:2011et}.
In the case, however, of many observatories \cite{obs1}-\cite{obs7}, especially as regards the older records, many data for interacting DE-DM exist; further, the decay of DE to the DM and unparticle is at best only an approximation \cite{Jamil:2011iu},\cite{Chen}, the success attending which probably varies considerably at different stations \cite{cr}. When the continuity equations are satisfied for different types of matter(energy) contents with interaction terms, a phase space description is found to be useful \cite{Jamil:2012nma}-\cite{Jamil:2012yz}. In this approach we present, a suitable multidimensional representation of a first order dynamical system  in which each dimension corresponds to one density of the system. Thus, a point of phase space corresponds to a specific state of the system, and a path represents the evolution of the density functions of the system through different states. Our aim in this letter is to perform phase space description for an interacting DE-DM system in a type of modified gravity, called as mimetic gravity (MG).

{\it Basic equations}
Another opportunity to solve DE problem will present a type of disformal invariant theory, where we might be identified with  a specific reparametrization of the metric $g_{\mu\nu}=(\tilde{g}^{\alpha\beta}
\partial_\alpha\phi\partial_\beta\phi)\tilde{g}_{\mu\nu}$  \cite{Chamseddine:2013kea}:
\begin{equation}\label{s}
S=-\frac{1}{2}\int d^4x\sqrt{-g(\tilde{g}_{\mu\nu,\phi})}\{\frac{R[g_{\mu\nu}(\tilde{g}_{\mu\nu},\phi)]}{\kappa^2}+\mathcal{L}_m\},
\end{equation}
Here we suppose that
 $\tilde{g}_{\mu\nu}$ represents an auxiliary metric and $\phi$ denotes an auxiliary (non ghost)scalar field. Such type of modified gravity recently has been studied from different points of view \cite{Chamseddine:2014vna}-\cite{Momeni:2015fea}. Derivation of the field equations are straightforward. If we vary the action (\ref{s}) with respect to the metric $g_{\mu\nu}$ and the scalar field we obtain the following set of field equations:

\begin{equation}\label{eqn1}
(G^{\mu\nu}-\kappa^2
T^{\mu\nu})-(G-\kappa^2T)g^{\mu\alpha}g^{\nu\beta}\partial_\alpha\phi\partial_\beta\phi=0,
\end{equation}
\begin{equation}\label{eqn2}
\frac{1}{\sqrt{-g}}\partial_\kappa[\sqrt{-g}(G-\kappa^2T)g^{\kappa\lambda}\partial_\lambda\phi]=\nabla_\kappa[(G-T)\partial^{\kappa}\phi]=0,
\end{equation}
The acceleration expansion of Universe in standard cosmology may be attributed to a homogenous and isotropic FLRW metric in the following form:
\begin{eqnarray}
&&ds^2=dt^2-a(t)^2dx^2\label{g}.
\end{eqnarray}
The set of FLRW Eqs. can be obtained using (\ref{eqn1},\ref{eqn2}) for metric (\ref{g}):
\begin{eqnarray}
&&3H^2=\kappa^2\rho_{eff}=\kappa^2(3p_{tot}-\frac{R}{\kappa^2})
\\&&2\dot{H}=-\kappa^2p_{eff}==\kappa^2(4p_{tot}-\frac{R}{\kappa^2})
\end{eqnarray}
here $p_{tot}=\Sigma_{i=1}^3 p_i=\Sigma_{i=1}^3 w_i\rho_i$.

We assume the matter contents are composed by three component fluid containing matter $\rho_m$, dark energy $\rho_d$ and
radiation $\rho_r$ . We suppose that these components are interacting with an unknown set of interction fuctions $\Gamma_i$. The corresponding continuity
equations for three components can be written in the following system:
\begin{eqnarray}\label{2a}
\dot \rho_d+3H(\rho_d+p_d)&=&\Gamma_1,\nonumber\\
\dot\rho_m+3H\rho_m&=&\Gamma_2,\\ \dot
\rho_r+3H(\rho_r+p_r)&=&\Gamma_3,\nonumber
\end{eqnarray}
To preserve the total continuty, the interaction functions must
 satisfy collectively such that
$\Gamma_1+\Gamma_2+\Gamma_3=0$. To define an autonomous dynamical system , it is needed to
define dimensionless density parameters via the following set of new variables:
\begin{equation}\label{3}
x\equiv\frac{\kappa^2\rho_d}{3H^2},\ \
y\equiv\frac{\kappa^2\rho_m}{3H^2},\ \
z\equiv\frac{\kappa^2\rho_r}{3H^2}.
\end{equation}
 Using the FLRW  Eqs we can calculate the following important quantity:
\begin{eqnarray}
&&\frac{\dot{H}}{H^2}=-3\Big(\Sigma_{a=1}^{3}w_ax_a+\frac{1}{2}\Big)\label{Hdot}.
\end{eqnarray}
here $x_a=(x,y,z)$. We change time coordinate from $t$ to the $N=\ln a$. Using (\ref{Hdot}),
we rewrite the continuity equations (\ref{2a}) in following autonomous system in the dimensionless variables:
\begin{eqnarray}\label{sys}
\frac{dx}{dN}&=&-3x\Big(w_d-2(w_dx+w_rz)\Big)+
\frac{\kappa^2\Gamma_1}{3H^3},\label{dyn1}\\
\frac{dy}{dN}&=&6y(w_dx+w_rz)+\frac{\kappa^2\Gamma_2}{3H^3},\label{dyn2}\\
\frac{dz}{dN}&=&-3z\Big(w_r-2(w_dx+w_rz)\Big)+
\frac{\kappa^2\Gamma_3}{3H^3}\label{dyn3},
\end{eqnarray}
 we shall assume $w_m=0$, $w_r=\frac{1}{3}$ and $-1\leq w_d<-\frac{1}{3}$.

 The coupling functions $\Gamma_i$, $i=1,2,3$ are considered as phenomenological general functions of
the energy densities $\rho_i$ and the Hubble parameter $H$ i.e.
$\Gamma_i(H\rho_i)$. If we forego this assumption, the form of interaction term is $\Gamma_i\sim \Sigma  b H \rho$ or $\Gamma_i\sim \Sigma H^{-1}\rho_a \rho_n$. In strictness the interaction term  $\Gamma$ must be supposed to act upon the coordinates (\ref{3}) in its actual condition, whereas in (\ref{dyn1}-\ref{dyn3}), previously cited, the system is supposed to be absolutely autonomous.

{\it Analysis of stability in phase space}
In this phase space approach it is assumed that the system (\ref{dyn1}-\ref{dyn3}) is autonomous, and this assumption is sufficiently accurate for any practical purpose to which the above systems would be applicable in the ordinary analysing of a local stability point. But this was based upon the assumption of a density-$H$ relation between interaction terms, the interaction function $\Gamma_i$ of which depended on their relative densities $\rho_i$ as well as on Hubble parameter $H$. This being the case, we are at liberty to make the assumption that the interaction function of each term in (\ref{dyn1}-\ref{dyn3}) (under specified conditions) is known, without thereby introducing any risk of self-contradiction in mathematical calculations. This assumption has the great advantage, that the last interaction terms in (\ref{dyn1}-\ref{dyn3}) now appears as the autonomous term of the components $(x,y,z)$. Four models are specified for our dynamical system which we will study the local stability points using them \cite{Jamil:2012nma}.

{\it Interacting model - I}
The first model
has the following interaction terms:
\begin{equation}\label{8a}
\Gamma_1=-6bH\rho_d, \ \ \Gamma_2=\Gamma_3=3bH\rho_d,
\end{equation}
The model to be considered in a scenario in which dark energy will decay into matter and radiation. The parameter $b>0$ represents decay rate.

Using (\ref{8a}), the system (\ref{dyn1}-\ref{dyn3}) we obtain:
\begin{eqnarray}\label{sys}
\frac{dx}{dN}&=&-3x\Big(w_d-2(w_dx+w_rz)\Big)-6bx\label{eqm1},\\\nonumber
\frac{dy}{dN}&=&6y(w_dx+w_rz)+3bx,\label{eqm2}\\\nonumber
\frac{dz}{dN}&=&-3z\Big(w_r-2(w_dx+w_rz)\Big)+
3bx\label{eqm3},
\end{eqnarray}
The stationary or critical points are the solutions by equating the left hand sides of (\ref{eqm1}-\ref{eqm3}) to zero. That assumption would create the possibility of the existence of stable point, that is, of any critical point of the system.
The possible solutions and the eigenvalues$\lambda$  of the linearized system $\frac{d\vec{y}}{dN}=A\vec{y}$ near these points (i.e. when $\vec{X}=\vec{X}_c+\vec{Y},||\vec{Y}||\\ \ll ||\vec{X}_c||$), as the solutions for the vector equation $A\vec{V}=\lambda\vec{V}$,  are classified as the following:
\begin{itemize}
\item $A_1$: $\vec{X}_c=( 0, 0, 0)$,  $\lambda_i=\{0,-3/2\,w_{{r}}-3/2\,w_{{d}}-3\,b-3/2|-2\,b-w_{{d}}+w_{{
r}} |
, -3/2\,w_{{r}}-3/2\,w_{{d}}-3\,b+3/2| -2\,b-w_{{d}}+w_{{
r}} |
\}$

\item  $B_1$:$\vec{X}_c=( 0, 0, 1/2)$, $\lambda_i=\{1,1-3/2\,w_{{d}}-3\,b-3/2\,| w_{{d}}+2\,b |,1-3/2\,w_{{d}}-3\,b+3/2| w_{{d}}+2\,b |\}$.

\item  $C_1$: \begin{eqnarray}
&&x_c=\,{\frac {-w_{{r}}w_{{d}}+{w_{{d}}}^{2}+4\,bw_{{d}}-2\,bw_{{r}}+4
\,{b}^{2}}{2(-w_{{r}}w_{{d}}+{w_{{d}}}^{2}+2\,bw_{{d}}-bw_{{r}})}}
\\&&
y_c=-\,{\frac {b \left( 2\,b-w_{{r}}+w_{{d}} \right) }{2(-w_{{r}}w_{{d}
}+{w_{{d}}}^{2}+2\,bw_{{d}}-bw_{{r}})}}
\\&&
z_c=-\,{\frac {b \left( w_{{d}}+2\,b \right) }{2(-w_{{r}}w_{{d}}+{w_{{d
}}}^{2}+2\,bw_{{d}}-bw_{{r}})}}
\end{eqnarray}

 $\lambda_i=
\{3w_d+6b,3w_d+6b,6\,b+3\,w_{{d}}-1\}$. The following theorem helped us to classify the critical points:\\
{\emph Theorem } : The point $\vec{X}_c$ is asymptotically stable for the system (\ref{dyn1}-\ref{dyn3})  if the real part of every eigenvalue is negative. It is unstable if any eigenvalue has a positive real part.\\
We conclude that the points $A_1,B_1$ are unstable, while the point
 $C_1$ can be a  stable (unstable) degenerate node.
\end{itemize}

\begin{figure}
\centering
 \includegraphics[scale=0.4] {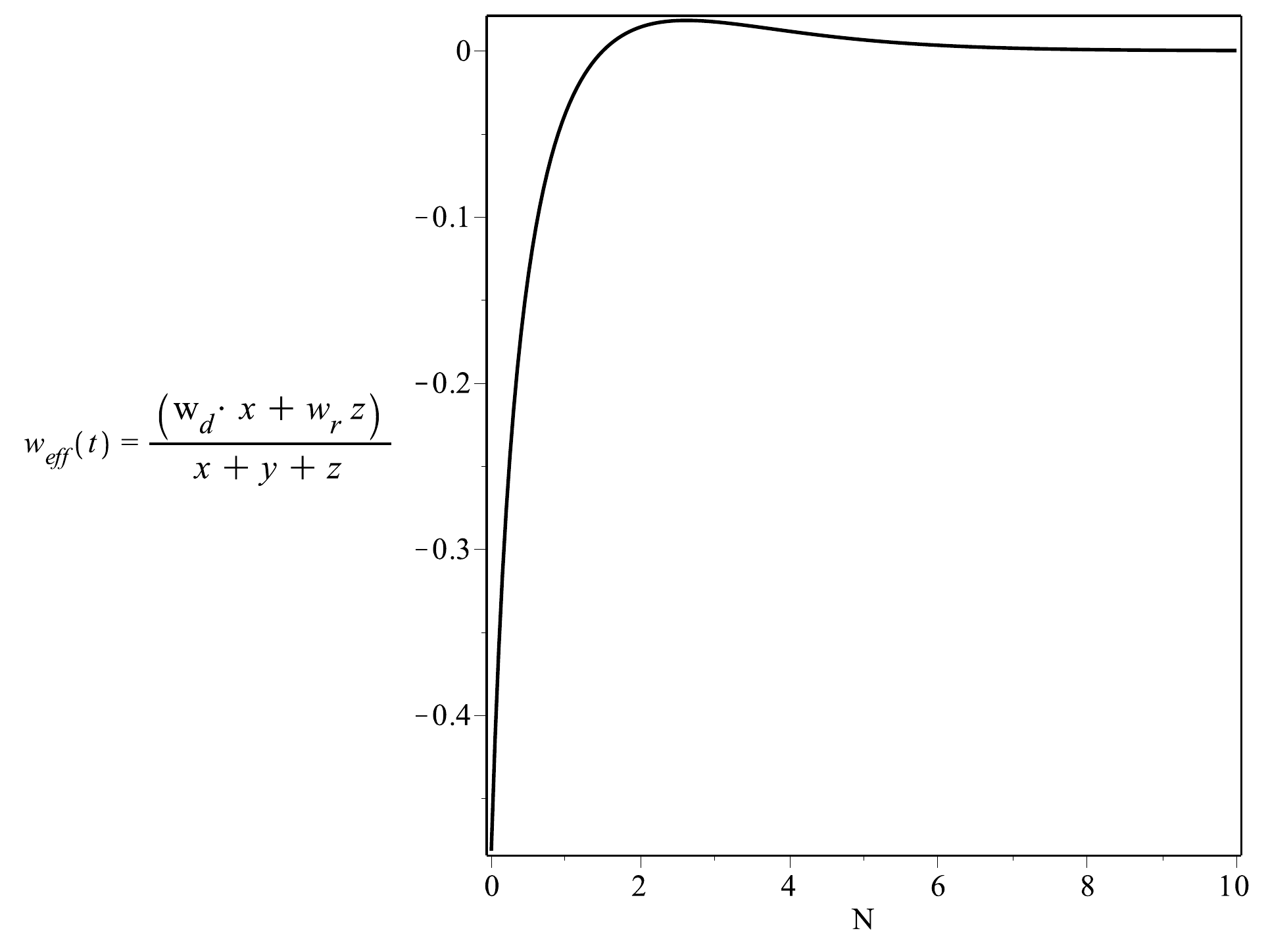}
  \caption{ Model I: Effective EoS $w_{eff}=\frac{w_d x+w_r z}{x+y+z}$ for $w_d=-0.7, b=0.5$.}
  \label{w1-eps-converted-to.pdf}
\end{figure}

The density functions functions $(x,y,z)$, phase portrait and the effective EoS are shown numerically in the following Fig.(\ref{x-y-z1-eps-converted-to.pdf},\ref{xyz1-eps-converted-to.pdf},\ref{w1-eps-converted-to.pdf}), in which the  parameters  are taken as the $w_d=-0.7, b=0.5$; the  solutions for density functions  are those determined by numerical algorithms, and the effective EoS $w_{eff}$ follow by calculation in the manner explained for
(\ref{eqm1}-\ref{eqm3}).

\begin{figure}
\centering
 \includegraphics[scale=0.4] {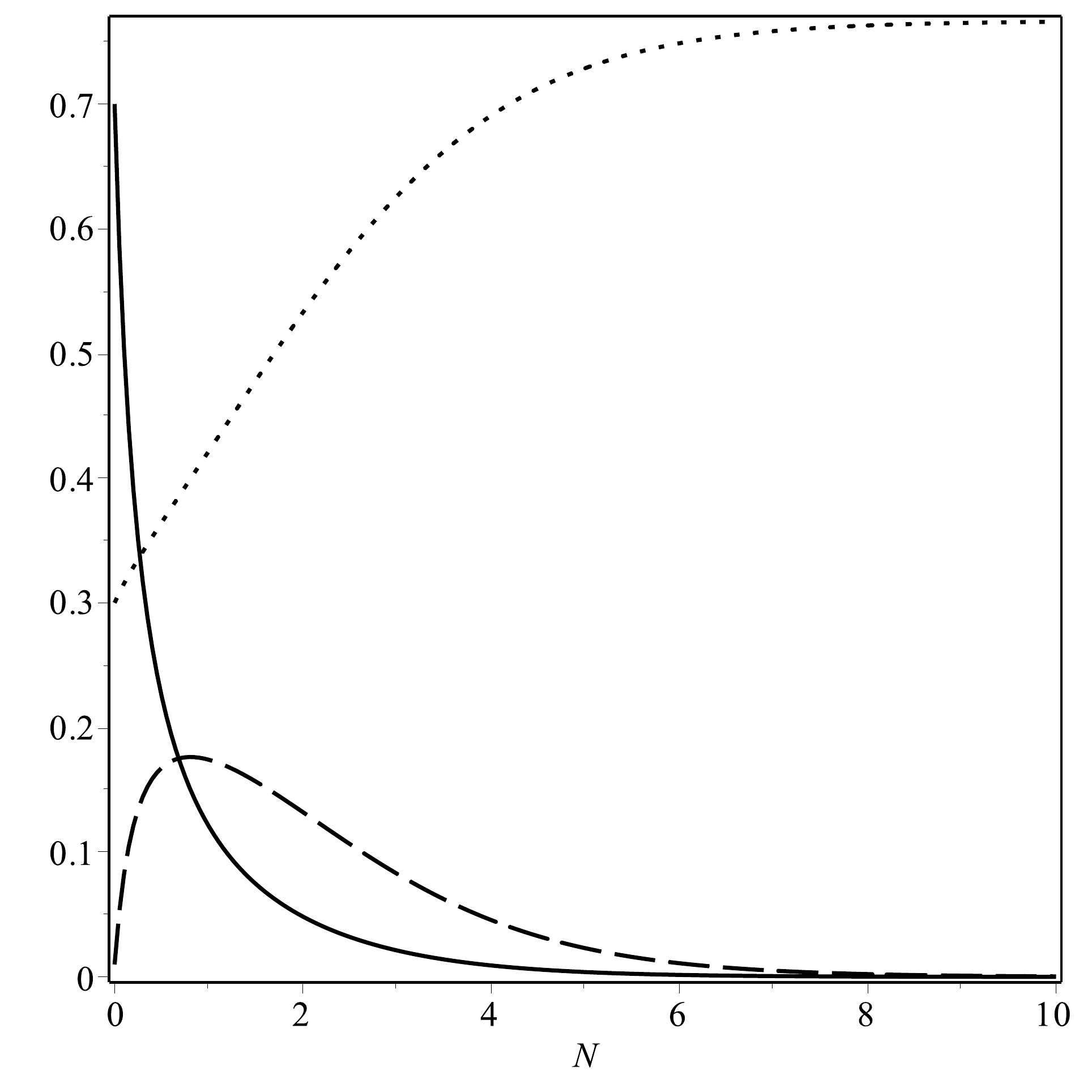}
  \caption{ Model I: Solutions $x(N),y(N),z(N)$ for $w_d=-0.7, b=0.5$. Solid $x(N)$, dot $y(N)$ and dash $z(N)$}
  \label{x-y-z1-eps-converted-to.pdf}
\end{figure}

\begin{figure}
\centering
 \includegraphics[scale=0.7] {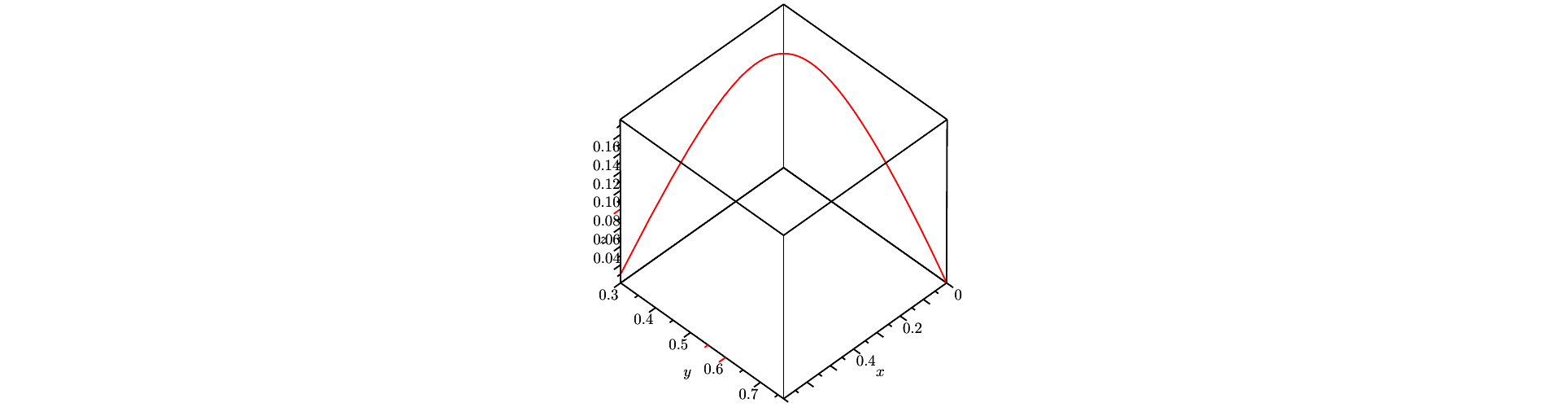}
  \caption{ Model I: Phase space for $w_d=-0.7, b=0.5$. It shows an attractor behavior. }
  \label{xyz1-eps-converted-to.pdf}
\end{figure}

{\it Interacting model - II} The second model of interaction is given by the following terms:

\begin{equation}\label{9a}
\Gamma_1=-3bH\rho_d,\ \ \Gamma_2=3bH(\rho_d-\rho_m),\ \
\Gamma_3=3bH\rho_m.
\end{equation}
The DE was a loss anyway, but the radiation  would be increased. Field equations can be written in the following forms:

\begin{eqnarray}\label{model2}
\frac{dx}{dN}&=&-3x\Big(w_d-2(w_dx+w_rz)\Big)-3bx\label{eqm1},\\\nonumber
\frac{dy}{dN}&=&6y(w_dx+w_rz)+3b(x-y),\label{eqm2}\\\nonumber
\frac{dz}{dN}&=&-3z\Big(w_r-2(w_dx+w_rz)\Big)+
3by\label{eqm3},
\end{eqnarray}
When we solved the Eqs. $\frac{dx_a}{dN}=0$, we obtained all critical points, eigenvalues of the linearized system:
\begin{itemize}
\item $A_2$: $\{x = 0, y = 0, z = 0\}$, $\lambda_i=\{-1,-3b,-3(w_d+b)$.
It may be considered in the system as a stable point.

\item  $B_2$:  $\{x = 0, y = 0, z = 1/2\}$,$\lambda_i=\{1,1-3b,1-3(w_d+b)\}$.
This point is unstable.

\item  $C_2$:  $\{x = 0, y = -(3/2)b+1/2, z = (3/2)b\}$, $\lambda_i=\{3b,3b-1,-3w_d\}$.
This critical point is also unstable.

\item  $D_2$:
\begin{eqnarray}
&&x = -\frac{w_d-3w_d^2-3bw_d}{6w_d^2+2b-2w_d},\\&&\nonumber
 y = -\frac{-b+3bw_d+3b^2}{6w_d^2+2b-2w_d},\\&&\nonumber
 z = \frac{3b^2}{6w_d^2+2b-2w_d}.
\end{eqnarray}

 \begin{eqnarray}
\lambda_1=3w_d,\ \
\lambda_2=3(w_d+b),\ \
\lambda_3=-1+3(w_d+b).
\end{eqnarray}
This critical point can be considered as unstable if $w_d<-b$.

\end{itemize}

Numerical solution, is obtained by imposing suitable IC and parameters into (\ref{model2}) . In Fig.(\ref{x-y-z2.eps}) we see an increasing form for radiation while the DE is decreased as well as DM.

\begin{figure}
\centering
 \includegraphics[scale=0.4] {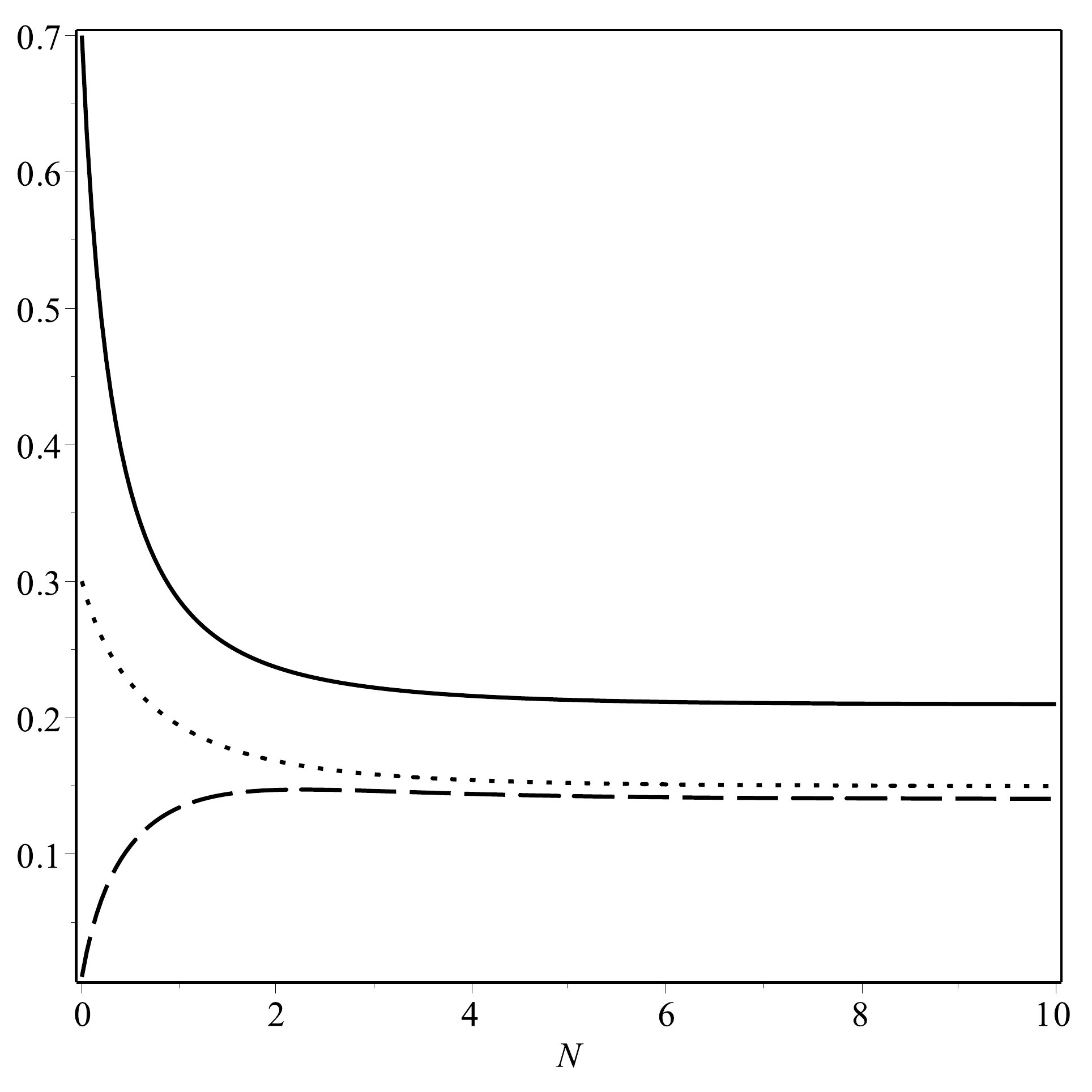}
  \caption{ Model II: Solutions $x(N),y(N),z(N)$ for $w_d=-0.7, b=0.5$. Solid $x(N)$, dot $y(N)$ and dash $z(N)$}
  \label{x-y-z2-eps-converted-to.pdf}
\end{figure}
An cosmological latetime attractor is detected in Fig. (\ref{xyz2.eps}).
\begin{figure}
\centering
 \includegraphics[scale=0.4] {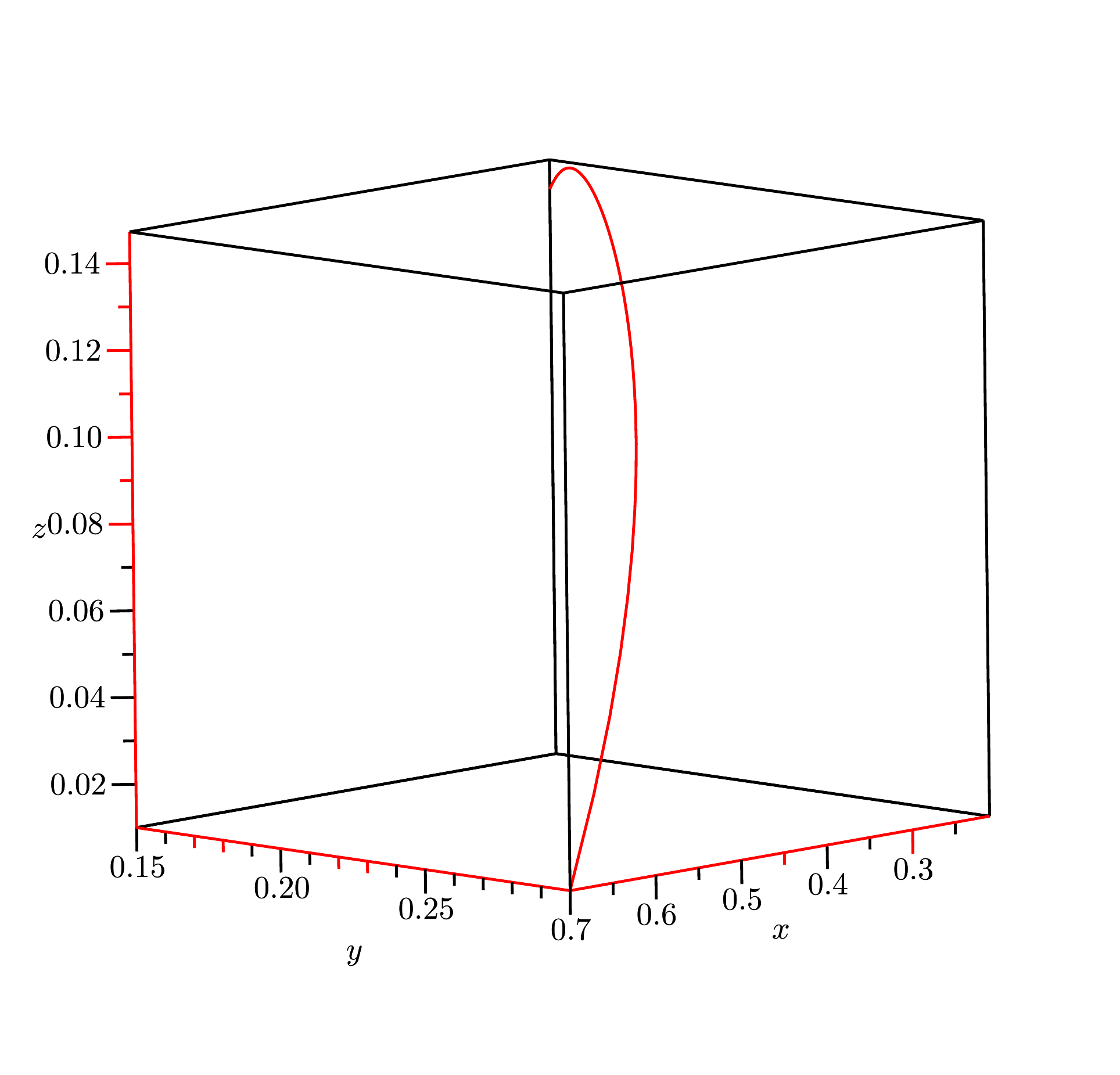}
  \caption{ Model II: Phase space for $w_d=-0.7, b=0.5$. It shows an attractor behavior. }
  \label{xyz2-eps-converted-to.pdf}
\end{figure}

Effective EoS is computed numerically. As we observe in Fig. (\ref{w2.eps}), the EoS $w_{eff}>-1$. So, the model can be considered as a cosmological ciable model for interacting DE-DM.
\begin{figure}
\centering
 \includegraphics[scale=0.4] {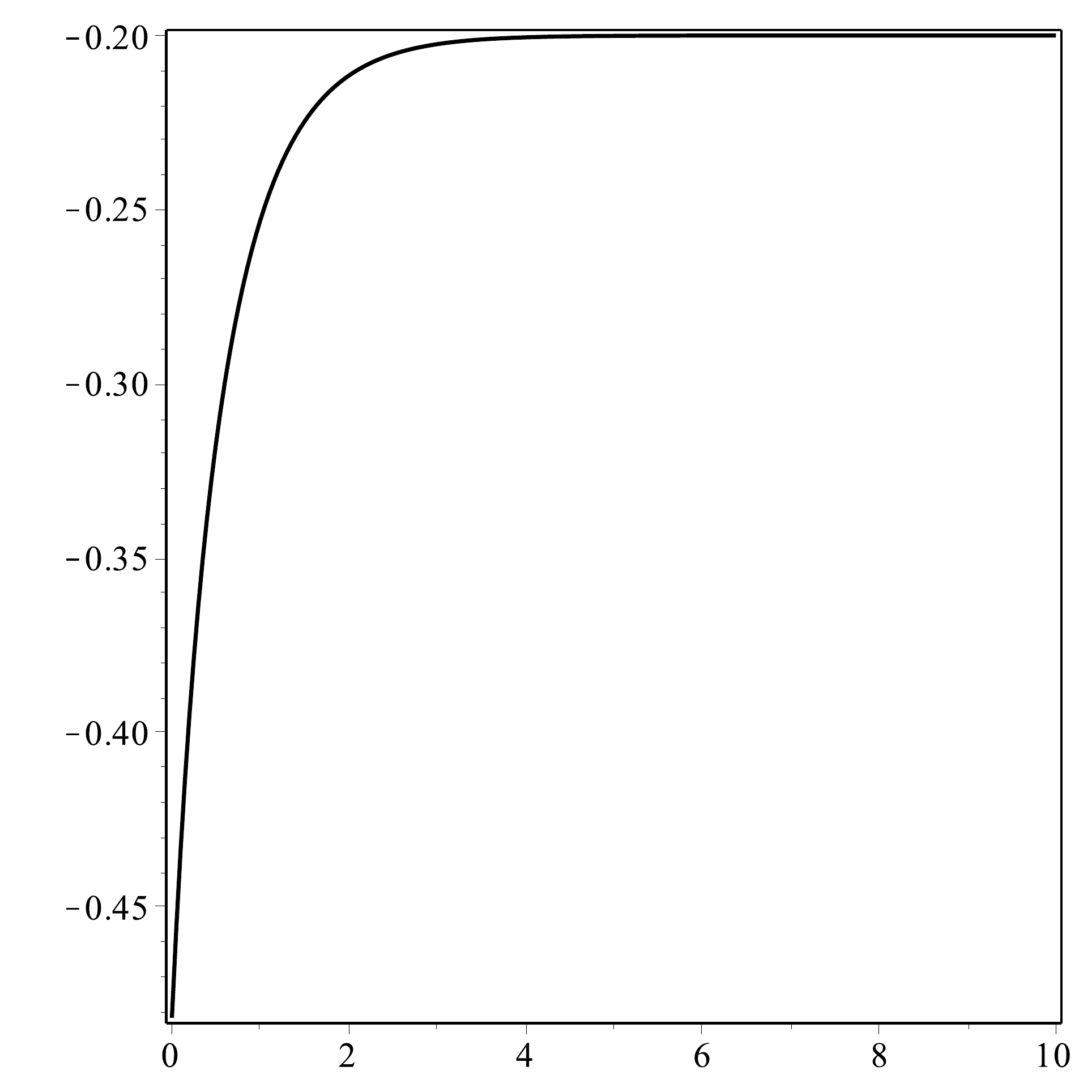}
  \caption{ Model II: Effective EoS $w_{eff}=\frac{w_d x+w_r z}{x+y+z}$ for $w_d=-0.7, b=0.5$.}
  \label{w2-eps-converted-to.pdf}
\end{figure}

{\it Interacting Model - III}
A third model which we'll consider is a scenario in which DE is decayed but DM and radiation are increasing:

\begin{equation}\label{12}
\Gamma_1=-6b\kappa^2 H^{-1}\rho_d\rho_r, \ \
\Gamma_2=\Gamma_3=3b\kappa^2 H^{-1}\rho_d\rho_r.
\end{equation}
The appropriate form of the dynamical system is taken as follows:
\begin{eqnarray}\label{model3}
\frac{dx}{dN}&=&-3x\Big(w_d-2(w_dx+w_rz)\Big)-18bxz\label{eqm1},\\\nonumber
\frac{dy}{dN}&=&6y(w_dx+w_rz)+9bxz,\label{eqm2}\\\nonumber
\frac{dz}{dN}&=&-3z\Big(w_r-2(w_dx+w_rz)\Big)+
9bxz\label{eqm3},
\end{eqnarray}

\begin{figure}
\centering
 \includegraphics[scale=0.4] {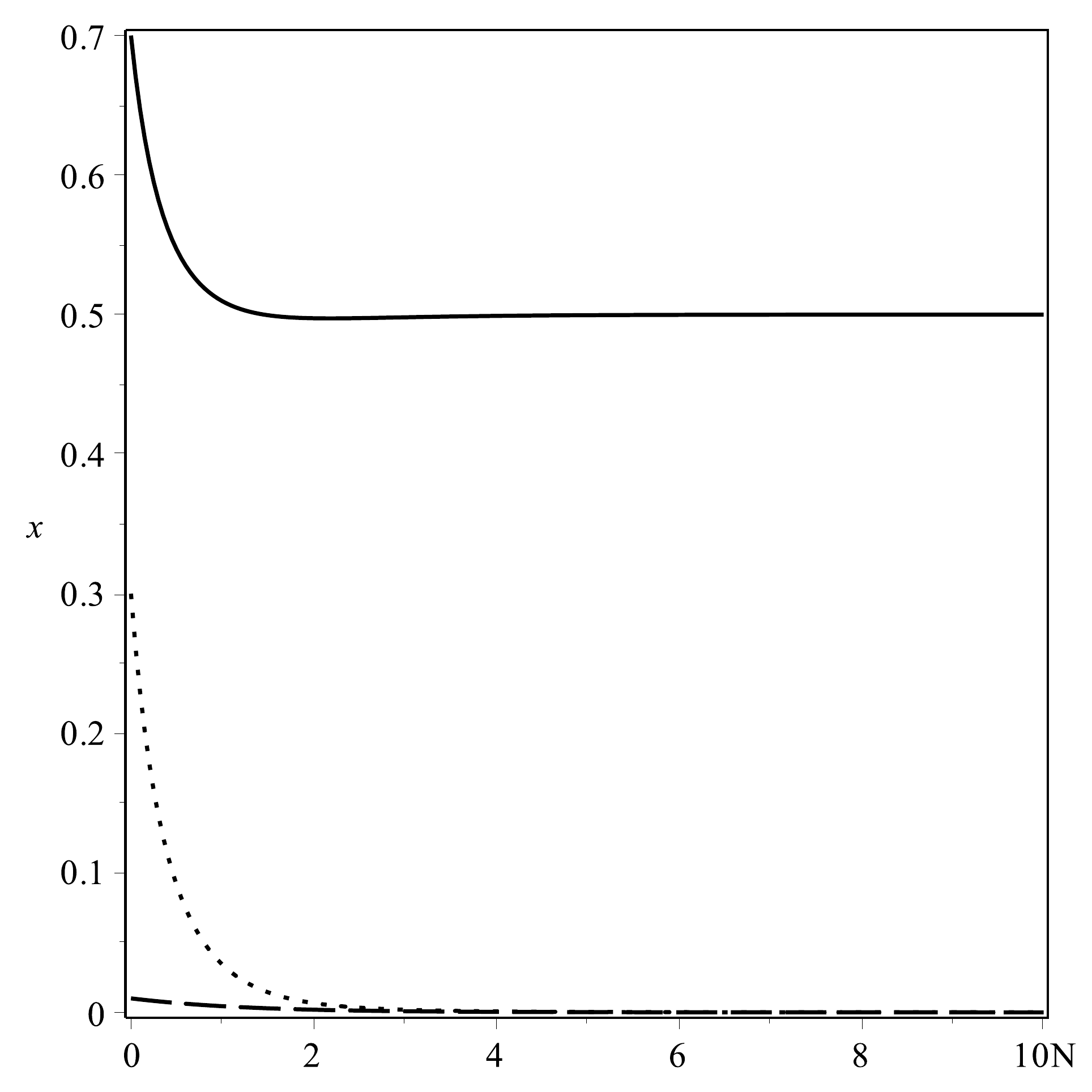}
  \caption{ Model III: Solutions $x(N),y(N),z(N)$ for $w_d=-0.7, b=0.5$. Solid $x(N)$, dot $y(N)$ and dash $z(N)$}
  \label{x-y-z3-eps-converted-to.pdf}
\end{figure}

The stationary points and the corresponding critical points for (\ref{model3}) are given as the following:
\begin{itemize}
\item $A_3$:$\{0,0,0\}$, $\lambda_i=\{0,-1,-3d\}$. This point is unstable.
\item $B_3$: $\{x = 0, y = 0, z = 1/2\}$,$\lambda_i=\{-3w_d+1-9b,1,1\}$. This point is also unstable.
\item $C_3$: $\{x = 1/2, y = 0, z = 0\}$,$\lambda_i=\{-1+3w_d+(9/2)b,3w_d,3w_d\}$. This critical point can be stable if $-1+3w_d+(9/2)b<0$.
\item $D_3$:
\begin{eqnarray}
&&x = -\frac{-3w_d+1-9b}{81b^2+54bw_d-9b}, \\&&\nonumber y = -\frac{-18w_d^2-81bw_d-2+12w_d-81b^2+27b}{162b^2+108bw_d-18b},\\&&\nonumber  z = -\frac{-2w_d+6w_d^2+9bw_d}{54b^2-6b+36bw_d}.
\end{eqnarray}
and eigenvalues:
\begin{eqnarray}
&&\lambda_1=\frac{3 w_d}{9 b+6 w_d-1},\\&&
\lambda_2=\,{\frac {9\,bw_{{d}}+\sqrt {3}\Delta}{6b \left( 9\,b+6\,w_{{d}}-1 \right) }}
\\&&
\lambda_3=\,{\frac {9\,bw_{{d}}-\sqrt {3}\Delta}{6b \left( 9\,b+6\,w_{{d}}-1 \right) }}.
\end{eqnarray}
here
\begin{eqnarray}
&&\frac{\Delta^2}{bw_{{d}}}=  -1377\,bw_{{d
}}+4860\,{b}^{2}w_{{d}}+2592\,{w_{{d}}}^{2}b\\&&\nonumber+432\,{w_{{d}}}^{3}+96\,w_
{{d}}-360\,{w_{{d}}}^{2}-1296\,{b}^{2}+180\,b-8+2916\,{b}^{3} .
\end{eqnarray}
There is (un)stablity for this critical point.
\end{itemize}
In Fig. (\ref{x-y-z3-eps-converted-to.pdf}) a numerical solution is developed for density function(s). All types of the densities are decaying.

\begin{figure}
\centering
 \includegraphics[scale=0.4] {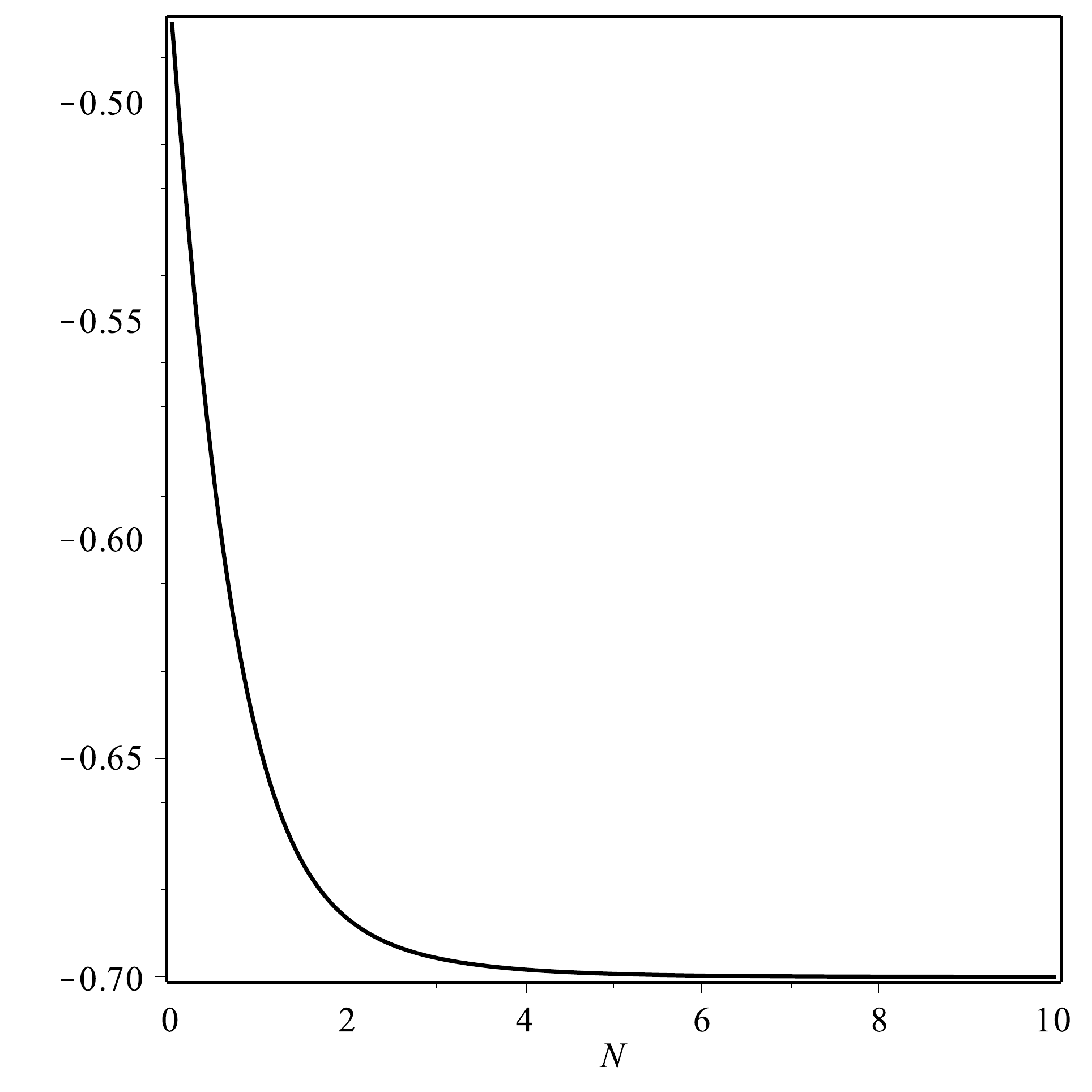}
  \caption{ Model III: Effective EoS $w_{eff}=\frac{w_d x+w_r z}{x+y+z}$ for $w_d=-0.7, b=0.5$.}
  \label{w3-eps-converted-to.pdf}
\end{figure}

An attractor which was started from an initial point is observed in Fig. (\ref{xyz3.eps}).
\begin{figure}
\centering
 \includegraphics[scale=0.4] {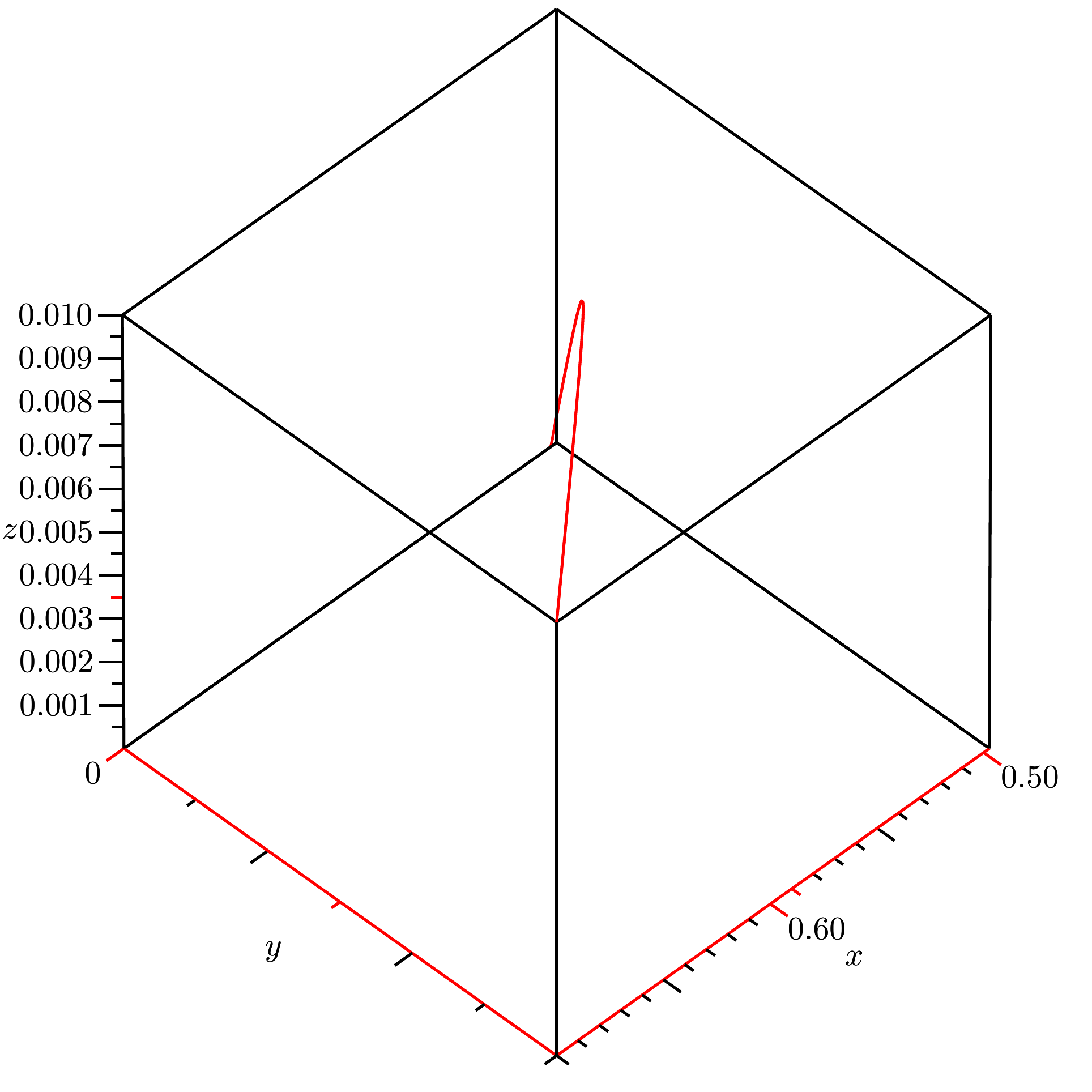}
  \caption{ Model III: Phase space for $w_d=-0.7, b=0.5$. It shows an attractor behavior. }
  \label{xyz3-eps-converted-to.pdf}
\end{figure}
An affective EoS with range $w_{eff}\leq -0.5$ is drawn in Fig. (\ref{w3-eps-converted-to.pdf}).

{\it Interacting Model - IV}

\begin{figure}
\centering
 \includegraphics[scale=0.4] {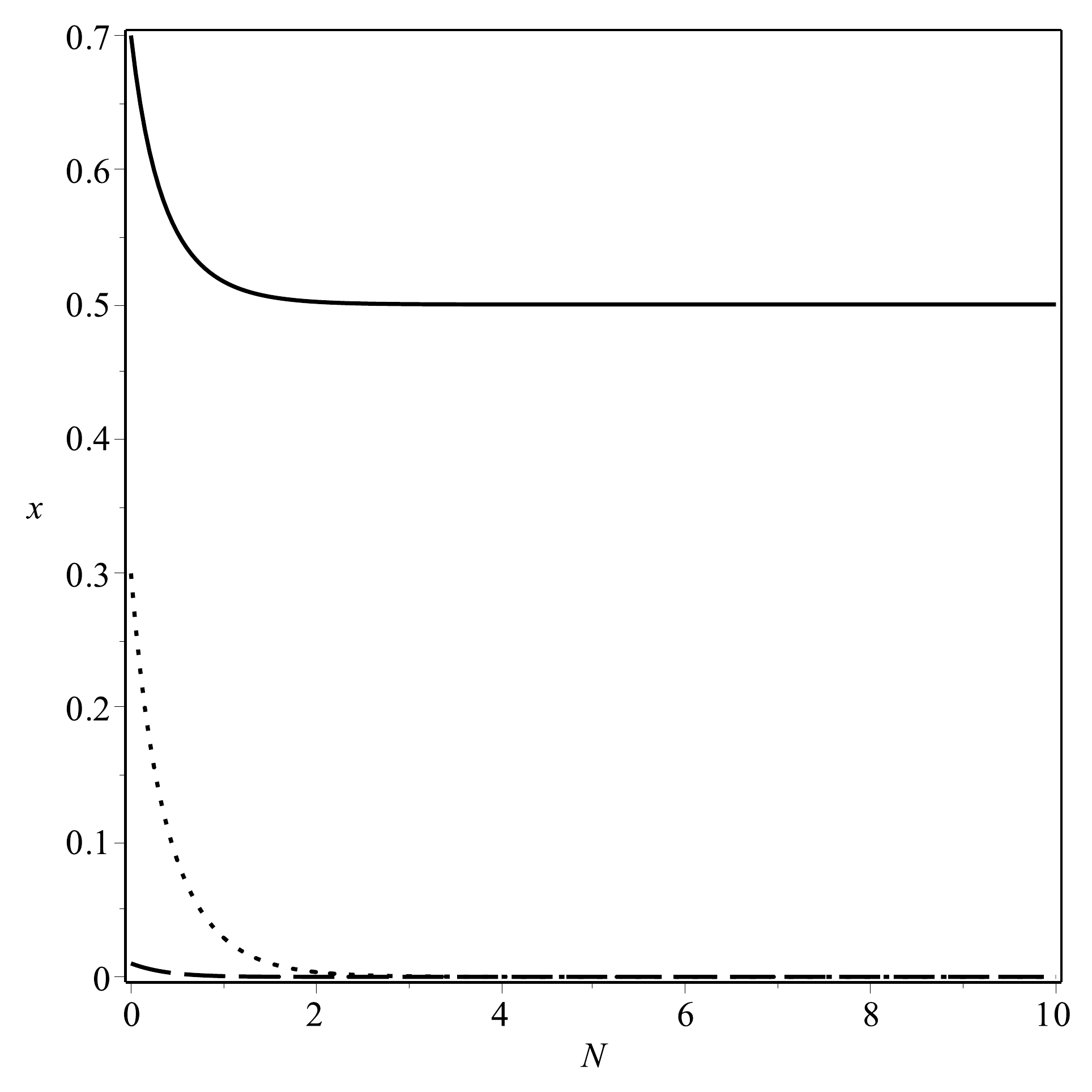}
  \caption{ Model IV: Solutions $x(N),y(N),z(N)$ for $w_d=-0.7, b=0.5$. Solid $x(N)$, dot $y(N)$ and dash $z(N)$}
  \label{x-y-z4.eps}
\end{figure}

An alternative form of interaction is defined by the following:
\begin{eqnarray}\label{11}
\Gamma_1&=&-3b\kappa^2 H^{-1}\rho_d\rho_r,\nonumber\\
\Gamma_2&=&3b\kappa^2 H^{-1}(\rho_d\rho_r-\rho_m\rho_r),\nonumber\\
\Gamma_3&=&3b\kappa^2 H^{-1}\rho_m\rho_r.
\end{eqnarray}
The corresponding dynamical system is written in the following form:
\begin{eqnarray}\label{sys}
\frac{dx}{dN}&=&-3x\Big(w_d-2(w_dx+w_rz)\Big)-9bxz\label{eqm1},\\\nonumber
\frac{dy}{dN}&=&6y(w_dx+w_rz)+9bz(x-y),\label{eqm2}\\\nonumber
\frac{dz}{dN}&=&-3z\Big(w_r-2(w_dx+w_rz)\Big)+
9byz\label{eqm3},
\end{eqnarray}
Critical points and the eigen values are obtained as the following:
\begin{itemize}
\item $A_4$: $\{x = 0, y = y, z = 0\}$, $\lambda_i=\{0,-1,-3w_d\}$. This is an unstable point.
\item $B_4$: $\{x = 0, y = 0, z = 1/2\}$, $\lambda_i=\{1,1-\frac{9b}{2},1-\frac{9b}{2}-3w_d\}$. This point is also unstable.
\item $C_4$: $\{x = 1/2, y = 0, z = 0\}$,$\lambda_i=\{3w_d-1,3w_d,3w_d\}$. The point here is stable.
\item $D_4$:
\begin{eqnarray}
&&x = \frac{1}{9b},\ \  y = \frac{1}{9b}
,\ \  z = -\frac{w_d}{3b}
\end{eqnarray}
we have a cubic equation for determining eigenvalues as a function of the absolute cosmological parameters $\{b,w_d\}$:
\begin{eqnarray}
&&{\lambda}^{3}-3\,w_{{d}}{\lambda}^{2}
-\frac{2w_{{d}} \left( -1+3
\,w_{{d}} \right) }{3b}\lambda\\&&\nonumber
+{\frac {{w_{{d}}}^{2} \left( -4+9\,b+6
\,w_{{d}} \right) }{b}}
=0
\end{eqnarray}
This cubic Eq. can be solved by quadratures. For $b>0$, $w_d<0$, it is possible to numerically show that one eigenvalue always is positive $\lambda_3>0$. So the system is unstable.

\item $E_4$:

\begin{eqnarray}
&&x = -\frac{1}{18}\frac{2-9b-6w_d}{b},\\&&\nonumber y = -\frac{27w_db+18w_d^2-9b-12w_d+2}{-18b+81b^2},\\&&\nonumber z = -\frac{-6w_d^2+2w_d}{-6b+27b^2}.
\end{eqnarray}
eigen values:
\begin{eqnarray}
&&{\lambda}^{3}-\,{\frac {18w_{{d}} \left( -1+3\,b+w_{{d}} \right) }{-2+9\,b}}{
\lambda}^{2}+A \lambda+B=0.
\end{eqnarray}

here
\begin{eqnarray}
&&3Aw_{{d}}^{-1}\left( -2+9\,b \right) ^{2}b=  -8-324\,{b}^{2}-72
\,{w_{{d}}}^{2}+72\,b\\&&\nonumber+48\,w_{{d}}-972\,w_{{d}}{b}^{2}y+4374\,w_{{d}}{b
}^{3}y+972\,{w_{{d}}}^{3}b+1458\,{w_{{d}}}^{2}{b}^{2}\\&&\nonumber-648\,{w_{{d}}}^{
2}b-486\,w_{{d}}{b}^{2}-1458\,{b}^{3}y+324\,{b}^{2}y+729\,{b}^{3}
\\&&
-B\left( -2+9\,b \right) ^{2}b\left( -1+3
\,w_{{d}} \right)^{-1} {w_{{d}}}^{-2}\\&&\nonumber=  -108\,b+8-216\,w_{{d}}by+972\,w
_{{d}}{b}^{2}y+216\,{w_{{d}}}^{3}\\&&\nonumber+324\,{w_{{d}}}^{2}b-486\,w_{{d}}{b}^
{2}\\&&\nonumber+1458\,{b}^{3}y-144\,{w_{{d}}}^{2}+108\,w_{{d}}b-648\,{b}^{2}y\\&&\nonumber+72\,
by-729\,{b}^{3}+486\,{b}^{2}
\end{eqnarray}
The eigenvalues has at least one positive real number $\lambda_{+}>0$. So the system is still unstable.

\end{itemize}
In Fig. (\ref{x-y-z4.eps}) a numerical solution is developed for density function(s). All types of the densities are decaying.

\begin{figure}
\centering
 \includegraphics[scale=0.4] {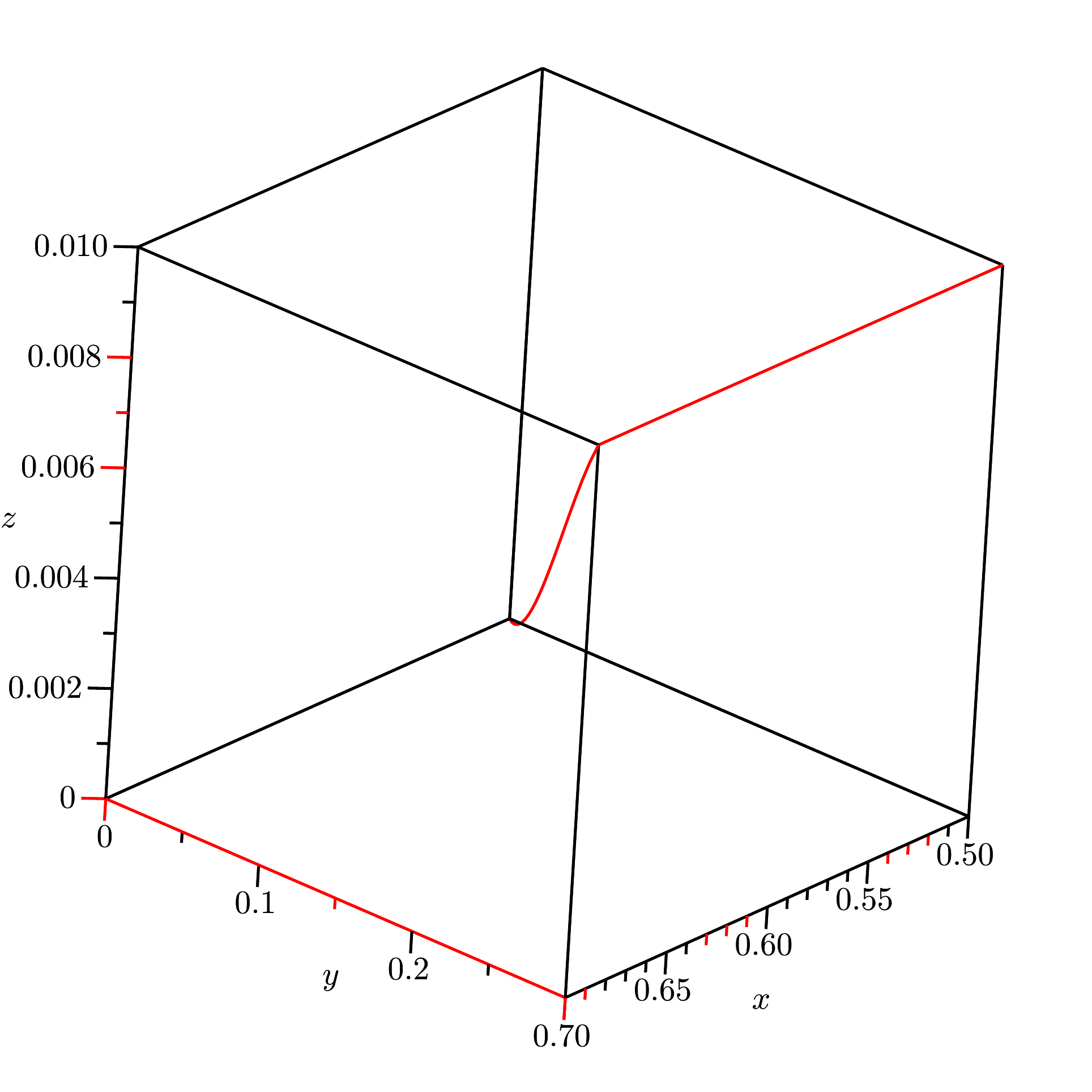}
  \caption{ Model IV: Phase space for $w_d=-0.7, b=0.5$. It shows an attractor behavior. }
  \label{xyz4-eps-converted-to.pdf}
\end{figure}

An attractor which was started from an initial point is observed in Fig. (\ref{xyz4-eps-converted-to.pdf}).

\begin{figure}
\centering
 \includegraphics[scale=0.4] {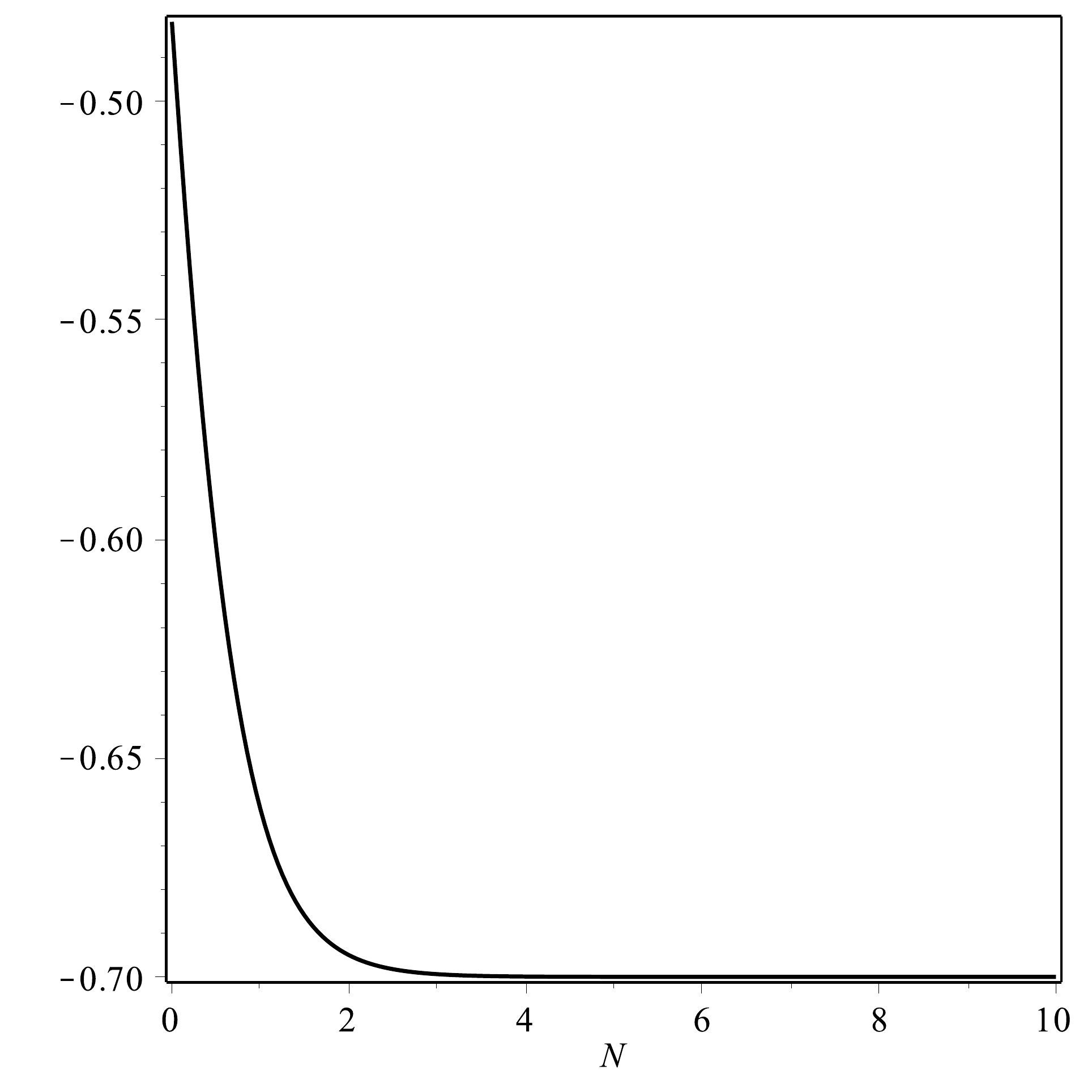}
  \caption{ Model IV: Effective EoS $w_{eff}=\frac{w_d x+w_r z}{x+y+z}$ for $w_d=-0.7, b=0.5$.}
  \label{w4-eps-converted-to.pdf}
\end{figure}

An affective EoS with range $w_{eff}\leq -0.5$ is drawn in Fig. (\ref{w4-eps-converted-to.pdf}).

\newpage
\section{Conclusion}
The final achievement of modified gravity in this direction was the extension of the method of the disformal transformations of the metric of arbitrary dimensions, successfully used by authors in the investigation of cosmological as well as of gravitational equalities, to any system whatever of mutually interacting components like dark energy and dark matter. We further supposed that the dark components are  interacting like fluids, but each perceiving what was passing in the other, and acting in consequence by interaction term $\Gamma_i$.
This letter was the first to investigate and describe in mimetic gravity the fact that a cosmological solution at a distance had the power of invariance  from initial condition to final state, and it also found that in some interacting cases the stable submanifold was produced.


\end{document}